\documentclass[%
reprint,
superscriptaddress,
showpacs,
 amsmath,amssymb,
 aps,
 prl,
]{revtex4-1}

\usepackage{graphicx}
\usepackage{dcolumn}
\usepackage{bm}
\usepackage{amsmath}
\usepackage{autobreak}
\usepackage{hyperref}
\usepackage[mathlines]{lineno}
\usepackage{natbib}
\usepackage{multirow}
\usepackage{subfigure}
\usepackage{float}

\begin{document}

\preprint{APS/123-QED}

\title{First experimental constraints on WIMP couplings in the effective field theory framework from CDEX}


\author{Y.~Wang}
\affiliation{Key Laboratory of Particle and Radiation Imaging (Ministry of Education) and Department of Engineering Physics, Tsinghua University, Beijing 100084}
\affiliation{Department of Physics, Tsinghua University, Beijing 100084}
\author{Z.~Zeng}
\affiliation{Key Laboratory of Particle and Radiation Imaging (Ministry of Education) and Department of Engineering Physics, Tsinghua University, Beijing 100084}
\author{Q. Yue}\altaffiliation [Corresponding author: ]{yueq@mail.tsinghua.edu.cn}
\affiliation{Key Laboratory of Particle and Radiation Imaging (Ministry of Education) and Department of Engineering Physics, Tsinghua University, Beijing 100084}
\author{L.~T.~Yang}\altaffiliation [Corresponding author: ]{yanglt@mail.tsinghua.edu.cn}
\affiliation{Key Laboratory of Particle and Radiation Imaging (Ministry of Education) and Department of Engineering Physics, Tsinghua University, Beijing 100084}
\author{K.~J.~Kang}
\affiliation{Key Laboratory of Particle and Radiation Imaging (Ministry of Education) and Department of Engineering Physics, Tsinghua University, Beijing 100084}
\author{Y.~J.~Li}
\affiliation{Key Laboratory of Particle and Radiation Imaging (Ministry of Education) and Department of Engineering Physics, Tsinghua University, Beijing 100084}

\author{M. Agartioglu}
\altaffiliation{Participating as a member of TEXONO Collaboration}
\affiliation{Institute of Physics, Academia Sinica, Taipei 11529}
\author{H.~P.~An}
\affiliation{Key Laboratory of Particle and Radiation Imaging (Ministry of Education) and Department of Engineering Physics, Tsinghua University, Beijing 100084}
\affiliation{Department of Physics, Tsinghua University, Beijing 100084}
\author{J.~P.~Chang}
\affiliation{NUCTECH Company, Beijing 100084}
\author{J.~H.~Chen}
\altaffiliation{Participating as a member of TEXONO Collaboration}
\affiliation{Institute of Physics, Academia Sinica, Taipei 11529}
\author{Y.~H.~Chen}
\affiliation{YaLong River Hydropower Development Company, Chengdu 610051}
\author{J.~P.~Cheng}
\affiliation{Key Laboratory of Particle and Radiation Imaging (Ministry of Education) and Department of Engineering Physics, Tsinghua University, Beijing 100084}
\affiliation{College of Nuclear Science and Technology, Beijing Normal University, Beijing 100875}

\author{C.~Y.~Chiang}
\altaffiliation{Participating as a member of TEXONO Collaboration}
\affiliation{Institute of Physics, Academia Sinica, Taipei 11529}

\author{W.~H.~Dai}
\affiliation{Key Laboratory of Particle and Radiation Imaging (Ministry of Education) and Department of Engineering Physics, Tsinghua University, Beijing 100084}

\author{Z.~Deng}
\affiliation{Key Laboratory of Particle and Radiation Imaging (Ministry of Education) and Department of Engineering Physics, Tsinghua University, Beijing 100084}
\author{X.~P.~Geng}
\affiliation{Key Laboratory of Particle and Radiation Imaging (Ministry of Education) and Department of Engineering Physics, Tsinghua University, Beijing 100084}
\author{H.~Gong}
\affiliation{Key Laboratory of Particle and Radiation Imaging (Ministry of Education) and Department of Engineering Physics, Tsinghua University, Beijing 100084}

\author{Q.~J.~Guo}
\affiliation{School of Physics, Peking University, Beijing 100871}

\author{X.~Y.~Guo}
\affiliation{YaLong River Hydropower Development Company, Chengdu 610051}

\author{H.~J.~He}
\affiliation{Key Laboratory of Particle and Radiation Imaging (Ministry of Education) and Department of Engineering Physics, Tsinghua University, Beijing 100084}
\affiliation{Department of Physics, Tsinghua University, Beijing 100084}

\author{L. He}
\affiliation{NUCTECH Company, Beijing 100084}
\author{S.~M.~He}
\affiliation{YaLong River Hydropower Development Company, Chengdu 610051}
\author{J.~W.~Hu}
\affiliation{Key Laboratory of Particle and Radiation Imaging (Ministry of Education) and Department of Engineering Physics, Tsinghua University, Beijing 100084}

\author{T.~S.~Huang}
\affiliation{Sino-French Institute of Nuclear and Technology, Sun Yet-sen University, Zhuhai 519082}

\author{H.~X.~Huang}
\affiliation{Department of Nuclear Physics, China Institute of Atomic Energy, Beijing 102413}

\author{H.~T.~Jia}
\affiliation{College of Physics, Sichuan University, Chengdu 610065}

\author{L.~P.~Jia}
\affiliation{Key Laboratory of Particle and Radiation Imaging (Ministry of Education) and Department of Engineering Physics, Tsinghua University, Beijing 100084}

\author{H.~B.~Li}
\altaffiliation{Participating as a member of TEXONO Collaboration}
\affiliation{Institute of Physics, Academia Sinica, Taipei 11529}


\author{J.~M.~Li}
\affiliation{Key Laboratory of Particle and Radiation Imaging (Ministry of Education) and Department of Engineering Physics, Tsinghua University, Beijing 100084}

\author{J.~Li}
\affiliation{Key Laboratory of Particle and Radiation Imaging (Ministry of Education) and Department of Engineering Physics, Tsinghua University, Beijing 100084}

\author{M.~X.~Li}
\affiliation{College of Physics, Sichuan University, Chengdu 610065}

\author{R.~M.~J.~Li}
\affiliation{College of Physics, Sichuan University, Chengdu 610065}

\author{X.~Li}
\affiliation{Department of Nuclear Physics, China Institute of Atomic Energy, Beijing 102413}

\author{Y.~L.~Li}
\affiliation{Key Laboratory of Particle and Radiation Imaging (Ministry of Education) and Department of Engineering Physics, Tsinghua University, Beijing 100084}
\author {B. Liao}
\affiliation{College of Nuclear Science and Technology, Beijing Normal University, Beijing 100875}
\author{F.~K.~Lin}
\altaffiliation{Participating as a member of TEXONO Collaboration}
\affiliation{Institute of Physics, Academia Sinica, Taipei 11529}
\author{S.~T.~Lin}
\affiliation{College of Physics, Sichuan University, Chengdu 610065}
\author{S.~K.~Liu}
\affiliation{College of Physics, Sichuan University, Chengdu 610065}
\author {Y.~D.~Liu}
\affiliation{College of Nuclear Science and Technology, Beijing Normal University, Beijing 100875}
\author {Y.~Y.~Liu}
\affiliation{College of Nuclear Science and Technology, Beijing Normal University, Beijing 100875}
\author{Z.~Z.~Liu}
\affiliation{Key Laboratory of Particle and Radiation Imaging (Ministry of Education) and Department of Engineering Physics, Tsinghua University, Beijing 100084}
\author{H.~Ma}
\affiliation{Key Laboratory of Particle and Radiation Imaging (Ministry of Education) and Department of Engineering Physics, Tsinghua University, Beijing 100084}

\author{Y.~C.~Mao}
\affiliation{School of Physics, Peking University, Beijing 100871}

\author{Q.~Y.~Nie}
\affiliation{Key Laboratory of Particle and Radiation Imaging (Ministry of Education) and Department of Engineering Physics, Tsinghua University, Beijing 100084}

\author{J.~H.~Ning}
\affiliation{YaLong River Hydropower Development Company, Chengdu 610051}
\author{H.~Pan}
\affiliation{NUCTECH Company, Beijing 100084}
\author{N.~C.~Qi}
\affiliation{YaLong River Hydropower Development Company, Chengdu 610051}

\author{C.~K.~Qiao}
\affiliation{College of Physics, Sichuan University, Chengdu 610065}

\author{J.~Ren}
\affiliation{Department of Nuclear Physics, China Institute of Atomic Energy, Beijing 102413}
\author{X.~C.~Ruan}
\affiliation{Department of Nuclear Physics, China Institute of Atomic Energy, Beijing 102413}


\author{C.~S.~Shang}
\affiliation{YaLong River Hydropower Development Company, Chengdu 610051}

\author{V.~Sharma}
\altaffiliation{Participating as a member of TEXONO Collaboration}
\affiliation{Institute of Physics, Academia Sinica, Taipei 11529}
\affiliation{Department of Physics, Banaras Hindu University, Varanasi 221005}
\author{Z.~She}
\affiliation{Key Laboratory of Particle and Radiation Imaging (Ministry of Education) and Department of Engineering Physics, Tsinghua University, Beijing 100084}

\author{L.~Singh}
\altaffiliation{Participating as a member of TEXONO Collaboration}
\affiliation{Institute of Physics, Academia Sinica, Taipei 11529}
\affiliation{Department of Physics, Banaras Hindu University, Varanasi 221005}
\author{M.~K.~Singh}
\altaffiliation{Participating as a member of TEXONO Collaboration}
\affiliation{Institute of Physics, Academia Sinica, Taipei 11529}
\affiliation{Department of Physics, Banaras Hindu University, Varanasi 221005}

\author {T.~X.~Sun}
\affiliation{College of Nuclear Science and Technology, Beijing Normal University, Beijing 100875}

\author{C.~J.~Tang}
\affiliation{College of Physics, Sichuan University, Chengdu 610065}
\author{W.~Y.~Tang}
\affiliation{Key Laboratory of Particle and Radiation Imaging (Ministry of Education) and Department of Engineering Physics, Tsinghua University, Beijing 100084}
\author{Y.~Tian}
\affiliation{Key Laboratory of Particle and Radiation Imaging (Ministry of Education) and Department of Engineering Physics, Tsinghua University, Beijing 100084}

\author {G.~F.~Wang}
\affiliation{College of Nuclear Science and Technology, Beijing Normal University, Beijing 100875}

\author{L.~Wang}
\affiliation{Department of Physics, Beijing Normal University, Beijing 100875}
\author{Q.~Wang}
\affiliation{Key Laboratory of Particle and Radiation Imaging (Ministry of Education) and Department of Engineering Physics, Tsinghua University, Beijing 100084}
\affiliation{Department of Physics, Tsinghua University, Beijing 100084}
\author{Y.~C.~Wang}
\affiliation{Key Laboratory of Particle and Radiation Imaging (Ministry of Education) and Department of Engineering Physics, Tsinghua University, Beijing 100084}
\author{Y.~X.~Wang}
\affiliation{School of Physics, Peking University, Beijing 100871}

\author{Z.~Wang}
\affiliation{College of Physics, Sichuan University, Chengdu 610065}

\author{H.~T.~Wong}
\altaffiliation{Participating as a member of TEXONO Collaboration}
\affiliation{Institute of Physics, Academia Sinica, Taipei 11529}
\author{S.~Y.~Wu}
\affiliation{YaLong River Hydropower Development Company, Chengdu 610051}
\author{Y.~C.~Wu}
\affiliation{Key Laboratory of Particle and Radiation Imaging (Ministry of Education) and Department of Engineering Physics, Tsinghua University, Beijing 100084}
\author{H.~Y.~Xing}
\affiliation{College of Physics, Sichuan University, Chengdu 610065}
\author{Y.~Xu}
\affiliation{School of Physics, Nankai University, Tianjin 300071}
\author{T.~Xue}
\affiliation{Key Laboratory of Particle and Radiation Imaging (Ministry of Education) and Department of Engineering Physics, Tsinghua University, Beijing 100084}

\author{Y.~L.~Yan}
\affiliation{College of Physics, Sichuan University, Chengdu 610065}

\author{N.~Yi}
\affiliation{Key Laboratory of Particle and Radiation Imaging (Ministry of Education) and Department of Engineering Physics, Tsinghua University, Beijing 100084}
\author{C.~X.~Yu}
\affiliation{School of Physics, Nankai University, Tianjin 300071}
\author{H.~J.~Yu}
\affiliation{NUCTECH Company, Beijing 100084}
\author{J.~F.~Yue}
\affiliation{YaLong River Hydropower Development Company, Chengdu 610051}
\author{M.~Zeng}
\affiliation{Key Laboratory of Particle and Radiation Imaging (Ministry of Education) and Department of Engineering Physics, Tsinghua University, Beijing 100084}

\author{B.~T.~Zhang}
\affiliation{Key Laboratory of Particle and Radiation Imaging (Ministry of Education) and Department of Engineering Physics, Tsinghua University, Beijing 100084}

\author{L.~Zhang}
\affiliation{College of Physics, Sichuan University, Chengdu 610065}

\author {F.~S.~Zhang}
\affiliation{College of Nuclear Science and Technology, Beijing Normal University, Beijing 100875}

\author{M.~G.~Zhao}
\affiliation{School of Physics, Nankai University, Tianjin 300071}
\author{J.~F.~Zhou}
\affiliation{YaLong River Hydropower Development Company, Chengdu 610051}
\author{Z.~Y.~Zhou}
\affiliation{Department of Nuclear Physics, China Institute of Atomic Energy, Beijing 102413}
\author{J.~J.~Zhu}
\affiliation{College of Physics, Sichuan University, Chengdu 610065}
\collaboration{CDEX Collaboration}
\noaffiliation

\date{\today}

\begin{abstract}We present weakly interacting massive particles (WIMPs) search results performed using two approaches of effective field theory from the China Dark Matter Experiment (CDEX), based on the data from both CDEX-1B and CDEX-10 stages. In the nonrelativistic effective field theory approach, both time-integrated and annual modulation analyses were used to set new limits for the coupling of WIMP-nucleon effective operators at 90\% confidence level (C.L.) and improve over the current bounds in the low $m_{\chi}$ region. In the chiral effective field theory approach, data from CDEX-10 were used to set an upper limit on WIMP-pion coupling at 90\% C.L. We for the first time extended the limit to the $m_{\chi}<$ 6 GeV/$c^2$ region.
\begin{description}
\item[PACS numbers]{95.35.+d,
29.40.-n,
98.70.Vc}
\end{description}
\end{abstract}

\maketitle


\section{Introduction}\label{section1}
Throughout past decades, compelling evidence from astroparticle physics and cosmology has indicated the existence of dark matter (DM) \cite{pdg_review}. The leading candidates for cold DM, weakly interacting massive particles (WIMPs, denoted as $\chi$) have been actively searched via spin-independent (SI) and spin-dependent (SD) elastic scattering with normal matter in underground direct detection experiments \cite{edelweiss3_dm_2016,lux_dm_2017,pico_2017,pandax_dm_2017,deap_2018,xmass_2019,cdex102018,cresst_light_2016,supercdms_2018,darkside_2018,xenon_light_2019}. However, the standard SI and SD scattering calculations are derived from the leading-order terms in WIMP-nucleon effective field theory (EFT) with ordinary treatment of nuclear structure \cite{wimp_lewin_1996,wimp_ressel,wimp_dimitrov}. To explore different possible WIMP-nucleus interactions, two alternative schemes of EFT, nonrelativistic effective field theory (NREFT) \cite{Fan_2010,Dobrescu_2006,Fitzpatrick_2013,Anand2014}, and chiral effective field theory (ChEFT) \cite{cheft_epelbaum,cheft_machleidt,cheft_hammer}have been proposed, and the consequences have been examined by several direct detection experiments \cite{supercdms_nreft_2015,xenon_nreft_2017,cresst_nreft_2019,pandax_eft_2019,lux_nreft}.

The China Dark Matter Experiment (CDEX) \cite{CDEX_introduction,cdex1,cdex12014,cdex12016,cdex0vbb2017,cdex1b2018,cdex102018,cdex_am,cdex_migdal,c10_darkphoton}, aiming at the direct detection of light DM with $p$-type point contact germanium (PPCGe) detectors, has finished two phases of data taking--namely CDEX-1(A, B) and CDEX-10. The energy threshold of 160 eV was achieved in CDEX-1B \cite{cdex1b2018} and CDEX-10 \cite{cdex102018}, which enhanced sensitivities for light DM. In this letter, we report the results of EFT analysis based on the data from CDEX-1B \cite{cdex1b2018,cdex_am} and CDEX-10 \cite{cdex102018,c10_darkphoton}. In addition, the long-duration data and stable running conditions of CDEX-1B\cite{cdex_am} allow annual modulation (AM) analysis to be performed as a new aspect of EFT studies.

\section{EFT approaches}\label{sec:2}
\subsection{NREFT framework}

The standard SI and SD analysis only includes the leading-order terms in WIMP-nucleon EFT, which may disappear or be suppressed in newer theories. In the NREFT approach, all leading-order and next-to-leading order operators maintaining Galilean-invariance are taken into consideration \cite{Fan_2010,Dobrescu_2006,Fitzpatrick_2013,Anand2014}. Because we are interested in elastic scattering direct detection, the amplitude can be calculated by modeling each scatter as a four-particle contact interaction with the Lagrangian density:
\begin{equation}
\begin{aligned}
  \mathcal{L}_{\text{int}}=\bar{\chi}\mathcal{O}_{\chi}\chi \bar{N}\mathcal{O}_{N}N \equiv \mathcal{O}\bar{\chi}\chi\bar{N}N,
\end{aligned}   
\end{equation} 
where $\chi$ and $N$ are nonrelativistic fields denoting the incident WIMP and target nucleon respectively. As shown in Eq. (2), this approach introduces 14 operators, $\mathcal{O}_{i} (i=1,3$-$15)$ \cite{Anand2014}, which rely on different types of nuclear responses. It is clear that operators $\mathcal{O}_1$ and $\mathcal{O}_4$ correspond to typical SI and SD interactions, respectively, which are considered in most direct detection. In some new models without typical interactions, new types of interactions become important, such as parity-violating interaction $i\overline{\chi}\gamma^{5}\chi\overline{N}N$, the nonrelativistic analog of which is $\mathcal{O}_{11}$. These operators explicitly depend on WIMP and nucleon spins, $\vec{S}_{\chi}$ and $\vec{S}_{N}$, relative perpendicular velocity between the WIMP and nucleon, $\vec{v}^{\bot}$, in addition to the momentum transfer $\vec{q}$. Of note, $\mathcal{O}_{2}$ is not considered here because it cannot be obtained from a relativistic operator at the leading order \cite{Fitzpatrick_2013,Anand2014}.     

\begin{align}
 &\mathcal{O}_{1}=1_{\chi}1_{N},   \:\:\:\:\:\:\:\:\:\:\:\:\:\:\:\:\:\:\:\:\:\:\:\:\:\:\:\:\:\:\:\:      \mathcal{O}_{9}=i\vec{S}_{\chi}\cdot[\vec{S}_{N}\times \frac{\vec{q}}{m_{N}}],\notag \\
 &\mathcal{O}_{3}=i\vec{S}_{N}\cdot[\frac{\vec{q}}{m_{N}}\times \vec{v}^{\bot}],  \:\:\:\:\:\:\:\:\:\:\:  \mathcal{O}_{10}=i\vec{S}_{N}\cdot \frac{\vec{q}}{m_{N}},\notag \\
 &\mathcal{O}_{4}=\vec{S}_{\chi}\cdot \vec{S}_{N},   \:\:\:\:\:\:\:\:\:\:\:\:\:\:\:\:\:\:\:\:\:\:\:\:\: \:\:   \mathcal{O}_{11}=i\vec{S}_{\chi}\cdot \frac{\vec{q}}{m_{N}},\notag \\
 &\mathcal{O}_{5}=i\vec{S}_{\chi}\cdot [\frac{\vec{q}}{m_{N}}\times \vec{v}^{\bot}],   \:\:\:\:\:\:\:\:\:\:\:\:  \mathcal{O}_{12}=\vec{S}_{\chi}\cdot [\vec{S}_{N}\times \vec{v}^{\bot}],\notag \\
 &\mathcal{O}_{6}=[\vec{S}_{\chi}\cdot \frac{\vec{q}}{m_{N}}]\cdot[\vec{S}_{N}\cdot \frac{\vec{q}}{m_{N}}],   \:\:  \mathcal{O}_{13}=i[\vec{S}_{\chi}\cdot \vec{v}^{\bot}][\vec{S}_{N}\cdot \frac{\vec{q}}{m_{N}}],\notag \\ 
 &\mathcal{O}_{7}=\vec{S}_{N}\cdot \vec{v}^{\bot},    \:\:\:\:\:\:\:\:\:\:\:\:\:\:\:\:\:\:\:\:\:\:\:\:\:\:\:\:\:      \mathcal{O}_{14}=i[\vec{S}_{\chi}\cdot \frac{\vec{q}}{m_{N}} ][\vec{S}_{N}\cdot \vec{v}^{\bot}],\notag \\
 &\mathcal{O}_{8}=\vec{S}_{\chi}\cdot \vec{v}^{\bot},  \:\:\:\:\:\:\:\:\:\:\:\:    \mathcal{O}_{15}=-[\vec{S}_{\chi}\cdot \frac{\vec{q}}{m_{N}} ][(\vec{S}_{N}\times \vec{v}^{\bot})\cdot \frac{\vec{q}}{m_{N}}]. \notag\\
\end{align} 

Each operator listed in Eq. (2) has distinct couplings $c_{i}$ to protons and neutrons with coefficients $c_{i}^{p}$ and $c_{i}^{n}$, respectively. Thus, the NREFT interaction takes the form:
\begin{equation}
\begin{aligned}
  \sum_{\alpha=n,p}\sum^{15}_{i=1}c_{i}^{\alpha}\mathcal{O}_{i}^{\alpha}, \:\: c_{2}^{\alpha}\equiv0.
\end{aligned}   
\end{equation} 

The differential scattering rate with respect to nuclear recoil energy in direct detection is generally given by:

\begin{equation}
\begin{aligned}
  \frac{dR}{dE_{R}}=\frac{\rho}{m_{\chi}m_{A}} \int_{v_{\text {min}}(E_R)}v{f(\bm{v})}\frac{d\sigma}{dE_{R}}d^3v,
\end{aligned}   
\end{equation} 
where $\rho=$ 0.3 GeV/($c^2$~cm$^{3}$) is the local DM density; $m_{\chi}$ and $m_A$ are the masses of the WIMP and target nucleus, respectively; $v$ is the relative velocity of WIMP in the lab-frame; $v_{\text{min}}$ is the minimal WIMP velocity that can induce a nuclear recoil for a given energy $E_{R}$, and $f(\bm{v})$ is a Maxwellian velocity distribution with a most probable speed of $v_{0} = 220$~ km/s and a galactic escape velocity of $v_{\text{esc}} = 544$~km/s.

\begin{figure}[H]
\centering\includegraphics[width=1.0\columnwidth]{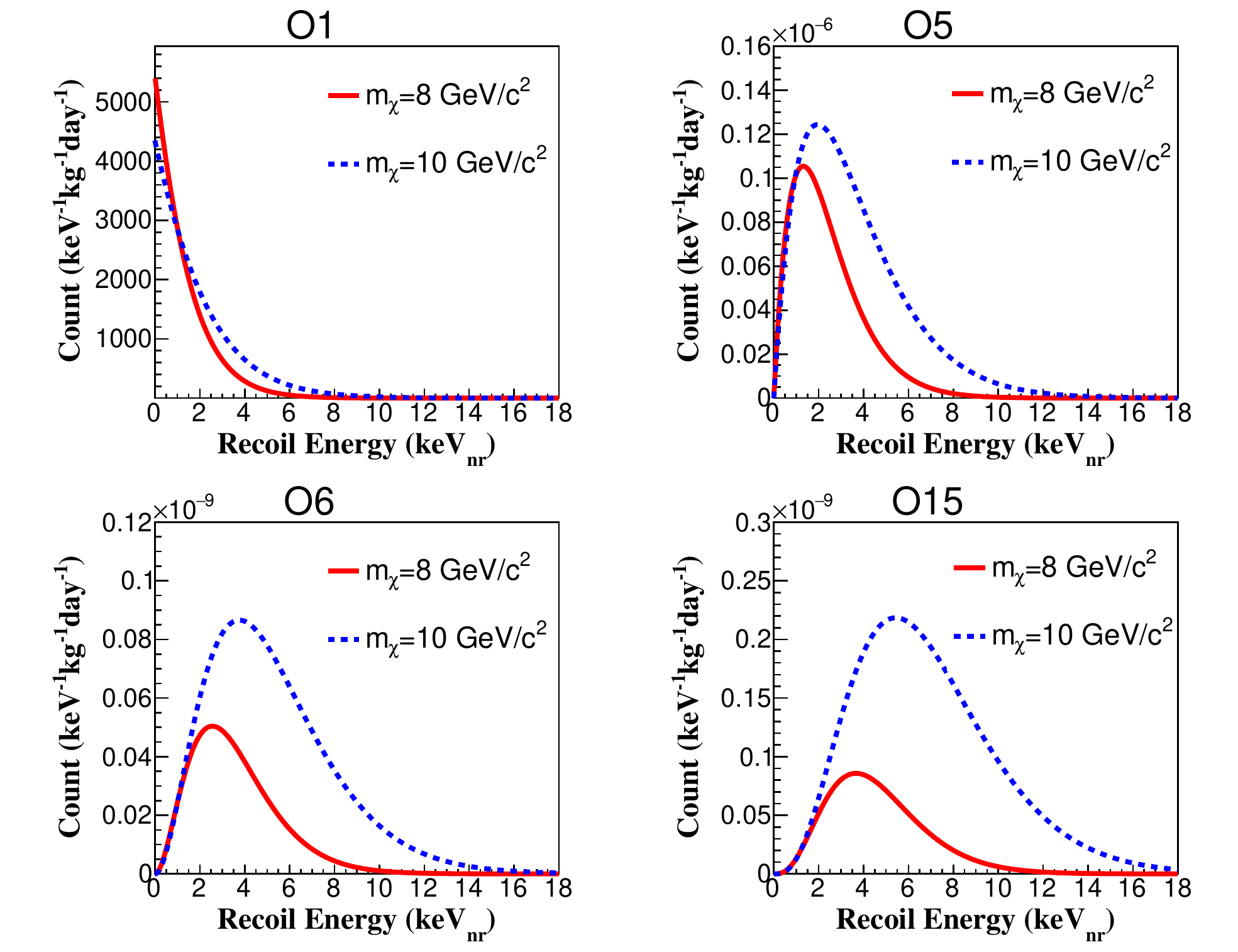}
\caption{Expected recoil spectra of isoscalar operator $\mathcal{O}_{1}$, $\mathcal{O}_{5}$, $\mathcal{O}_{6}$ and $\mathcal{O}_{15}$ for elastic scattering of 8 GeV/$c^{2}$ and 10 GeV/$c^{2}$ WIMPs on germanium nuclei, with $c^{0}_{i}=1$. The event rates of $\mathcal{O}_{5,6,15}$ are suppressed at low energies because they depend on the momentum transfer with the form of $\sim q^2$, $q^4$ and $q^6$ respectively.} 
\label{fig:1}
\end{figure}

As discussed in Ref. \cite{Anand2014}, the details of particle physics arising from NREFT are contained in the differential cross section $d\sigma/dE_{R}$, which is associated with 14 space-spin operators, that are distinctly coupled to protons and neutrons. Considering the isoscalar case, where the operators ($\mathcal{O}_{i}$) are equally coupled to protons and neutrons, the strength of the NREFT interaction is parameterized by coefficients $c^{0}_{i}=\frac{1}{2}(c_{i}^{p}+c_{i}^{n})$ (0 denotes the ``isoscalar" case), that have dimensions of 1/energy$^2$ such that they are multiplied by weak mass scale ($m_W$ =246.2 GeV/$c^2$) to be dimensionless. The results of NREFT framework presented in this letter are based on pure-isoscalar couplings. The expected recoil energy spectra of each WIMP mass for each EFT operator are calculated by a $Mathematica$ script introduced in Ref. \cite{Anand2014,Fitzpatrick_2013}. The spectra of $\mathcal{O}_{1}$, $\mathcal{O}_{5}$, $\mathcal{O}_{6}$ and $\mathcal{O}_{15}$ are shown in Fig.~\ref{fig:1}. The different dependence on momentum transfer of these operators results in different suppression of event rates at low energies in the corresponding spectra.

\begin{figure}[H]
\centering\includegraphics[width=1.0\columnwidth]{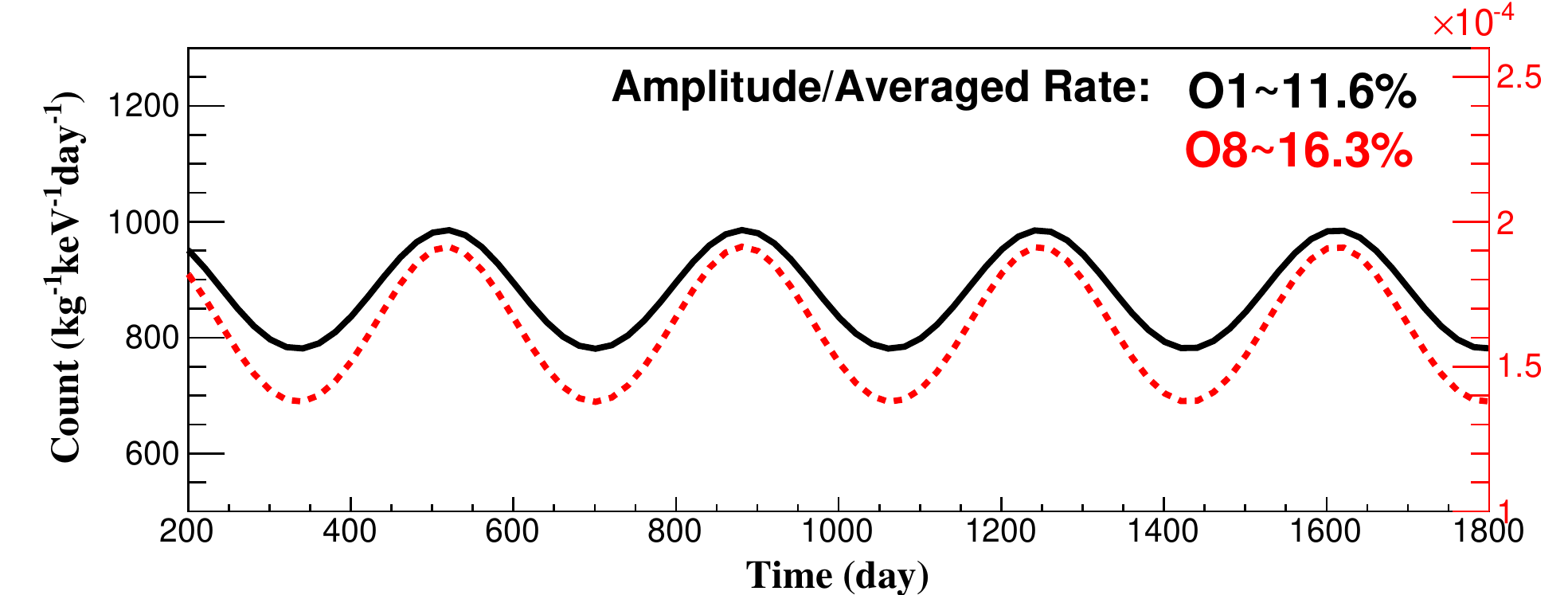}
\caption{Modulated event rates of $\mathcal{O}_{1}$ and $\mathcal{O}_{8}$ in the energy bin of 0.25$-$0.30 keV. The mass of WIMPs was set to 5 GeV/$c^2$, and $c^{0}_{1,8}=1$ .} 
\label{fig:2}
\end{figure}

It is known that Earth's velocity relative to the galactic WIMP halo is time-varying with
a period of one year and can be expressed as $ v_{E}(t)=\{232+30\times0.51\text{cos}{2\pi(t-\phi)}/{T}\}$~{km/s}, where $T$ is 365.25 days, and $\phi$ is 152.5 days from January 1$^{st}$. Positive observations of AM provide smoking-gun signatures for WIMP-nuclei scattering rates, as shown in Fig.~\ref{fig:2}, and the modulation amplitudes are proportional to isoscalar coefficients $({c^{0}_{i}})^2$. As discussed in Ref. \cite{Anand2014}, these operators are classified as leading order, next-to-leading order and next-to-next-to-leading order, depending on the total number of momenta and velocities involved. The cross sections of different operators are scaled by $(v^{2})^{\alpha}$, where $\alpha$ is the total number of momenta and velocities in these operators. Compared to $\mathcal{O}_{1}$, where the cross section is scaled by $v^{0}$, the modulation amplitude of $\mathcal{O}_{8}$ is larger because its cross section is scaled by $v^{2}$.

\subsection{ChEFT framework}
ChEFT is a low energy effective theory of quantum chromodynamics (QCD) that preserves QCD symmetries \cite{cheft_epelbaum,cheft_machleidt,cheft_hammer}. At low momentum scales on the order of pion mass, which is similar to the typical momentum transforms involved in WIMP scattering off nuclei, ChEFT predicts that pions can emerge as explicit degrees of freedom in addition to nucleons and reveals a new class of contributions referred to as two-body currents \cite{xenon_pion}. Such two-body contributions occur, e.g., when WIMP couples to a virtual pion that is exchanged between two nucleons \cite{cheft_epelbaum,cheft_machleidt,cheft_hammer}. The ChEFT approach can be mapped onto the single-nucleon couplings of NREFT, and the relation of these two approaches has been discussed in Ref. \cite{cheft_martin_4,cheft_martin}. In this study, the two frameworks are independently analyzed.

The two-body currents of the SI channel involve a new combination of hadronic matrix elements and Wilson coefficients and constitute the most important coherent corrections \cite{cheft_cirigliano,cheft_cirigliano_2,cheft_martin,cheft_martin_2,cheft_korber,cheft_martin3,cheft_andreoli,cheft_martin_4,collider_dm_2010}. Among these two-body currents, the scalar pion-exchange currents enter at the same order in the ChEFT power counting as momentum-suppressed single-nucleon currents, and new coefficient $c_{\pi}$ can give rise to WIMP scattering by pion-exchange between two nucleons \cite{xenon_pion,cheft_martin_4} with cross section denoted by $\sigma_{\chi\pi}^{\text{scalar}}$. In scenarios where the leading SI contribution vanishes, the scalar WIMP-pion scattering becomes important. Therefore, this study focuses on ChFET in the SI channel with only WIMP-pion coupling, which is similar to what is adopted in Ref. \cite{xenon_pion}. The inclusion of leading two-body currents in the SD channel is a correction to the standard response and is expected to have a considerable effect on SD search \cite{xenon100_sd_2013,xmass_sd_2014,xenon100_2016,pandax_sd_2017,lux_sd_2017}. However, the abovementioned study is beyond the scope of this work. 

With only scalar WIMP-pion coupling, the WIMP-nucleus ($\chi$-$\mathcal{N}$) cross sections can be expressed in terms of the scalar WIMP-pion cross section $\sigma_{\chi\pi}^{\text{scalar}}$ and WIMP-pion reduced mass $\mu_{\pi}$: 
\begin{equation}
\begin{aligned}
 \frac{d\sigma_{{\chi}{\mathcal N}}^{\text{SI}}}{dq^2}= \frac{\sigma_{{\chi}{\pi}}^{\text{scalar}}}{{\mu_{\pi}^2}v^2}|{\mathcal F}_{\pi}(q^2)|^2, \:\:\:\sigma_{{\chi}{N}}^{\text{scalar}}=\frac{\mu_{\pi}^2}{4\pi}|c_{\pi}|^2,
\end{aligned}   
\end{equation}
where $q$ is momentum transfer and $\mathcal{F}_{\pi}$ is the nuclear structure factor. The calculation of structure factor $\mathcal{F}_{\pi}$ has been discussed in detail in Ref. \cite{cheft_martin_4}. For a given WIMP mass $m_{\chi}$, the differential event rate for the WIMP-pion coupling can be written as
\begin{equation}
\begin{aligned}
\frac{dR}{dE_{R}}=\frac{2\rho{\sigma_{\chi\pi}^{\text{scalar}}}}{m_{\chi}\mu_{\pi}^2}\times|{\mathcal F}_{\pi}(q^2)|^2\times \int_{v_{\text {min}}(E_R)}\frac{f(\bm v)}{v}d^3v.
\end{aligned}   
\end{equation} 

\begin{figure}[H]
\centering\includegraphics[width=1.0\columnwidth]{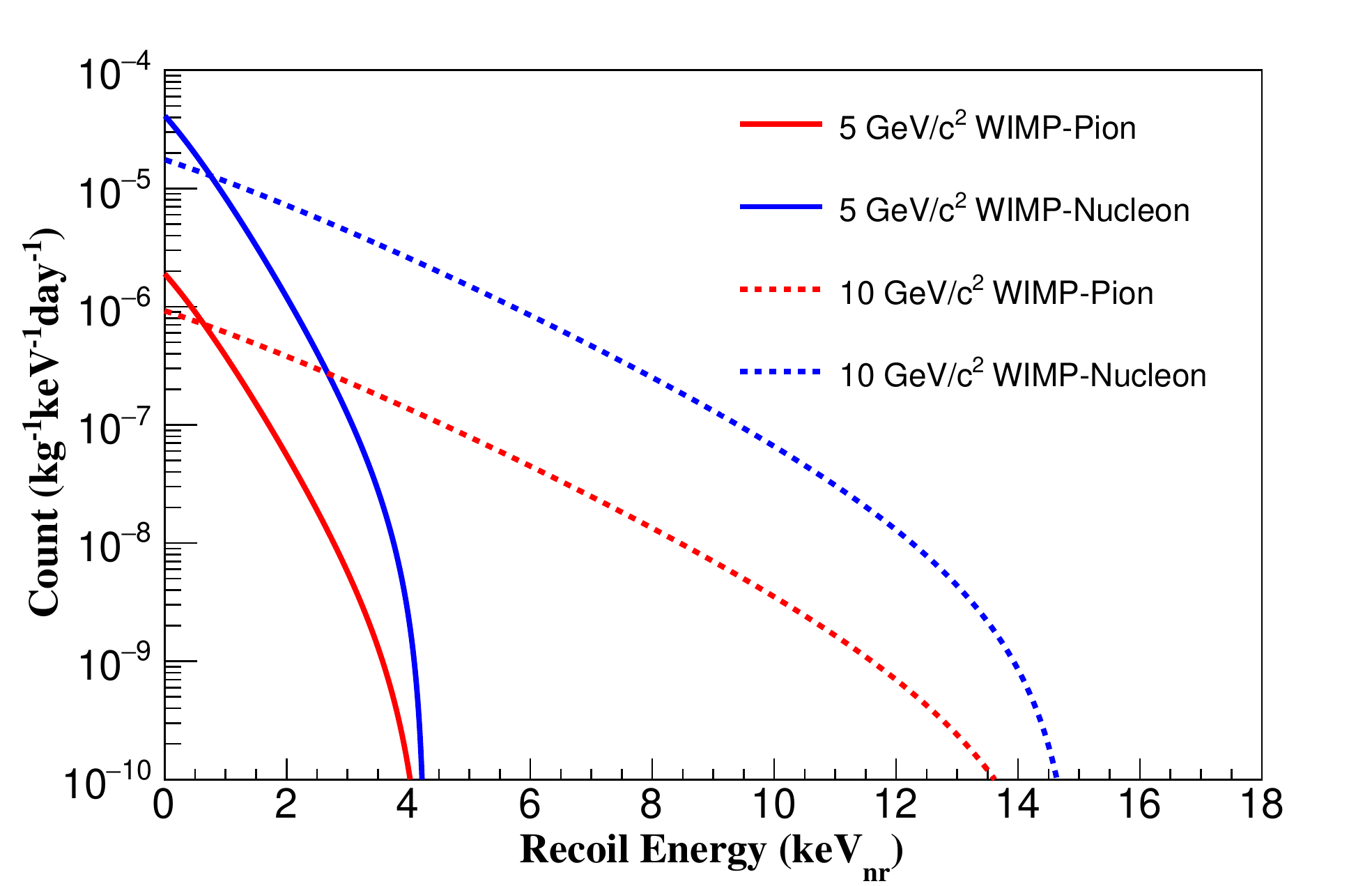}
\caption{Differential recoil spectrum for WIMP-pion (red) and SI WIMP-nucleon (blue) interactions at masses of 5 GeV/$c^2$ and 10 GeV/$c^2$ respectively, with the cross sections of $1\times10^{-46}$ cm$^{2}$.} 
\label{fig:3}
\end{figure}

The spectra of WIMP-pion scattering are shown in Fig.~\ref{fig:3}. Based on Eq. (6), we can derive the limits for $\sigma_{\chi\pi}^{\text{scalar}}$ as a function of the WIMP mass $m_{\chi}$. The coupling of WIMP to pion is estimated from nuclear structure factors. Through calculation, it is determined that the scaling of structure facture is approximately $(m_{\pi}/{\Lambda}_{\chi})^3A$, with nucleon number $A\sim70$ for germanium, pion mass $m_{\pi}\sim 135\:\text{MeV}$, and ChEFT breakdown scale $\Lambda_{\chi}\sim500-600\:\text{MeV}$. It is sub-leading compared to SI ($\mathcal{O}(A)$) but dominant compared to SD ($\mathcal{O}(1)$) scattering \cite{xenon_pion}.

\section{EXPERIMENT AND DATA ANALYSIS}\label{sec:3}
For the future ton-scale DM experiment, two generation experiments of CDEX have been designed for the direct detection of low-mass WIMPs with p-type point contact germanium detectors (PPCGe) at China Jinping Underground Laboratory \cite{cjpl}. In the first generation CDEX-1A and CDEX-1B experiments  \cite{cdex1,cdex12014,cdex12016,cdex1b2018}, 1-kg-scale single-element PPCGes cooled by a cold finger were used; while the second generation, CDEX-10 \cite{cdex102018,cdex10_tech}, was composed of three triple-element PPCGe strings that were directly immersed in liquid nitrogen. Owing to the problems of detector faults, seven elements of the CDEX-10 detector array are not working, while the other two in different strings named C10-B1 and C10-C1 are put into operation. The schematic diagrams of electronics and data acquisition (DAQ) system for CDEX-1B and CDEX-10 are shown in Ref. \cite{cdex1b2018,cdex10_tech}. 

\begin{figure}[H]
\centering\includegraphics[width=\linewidth]{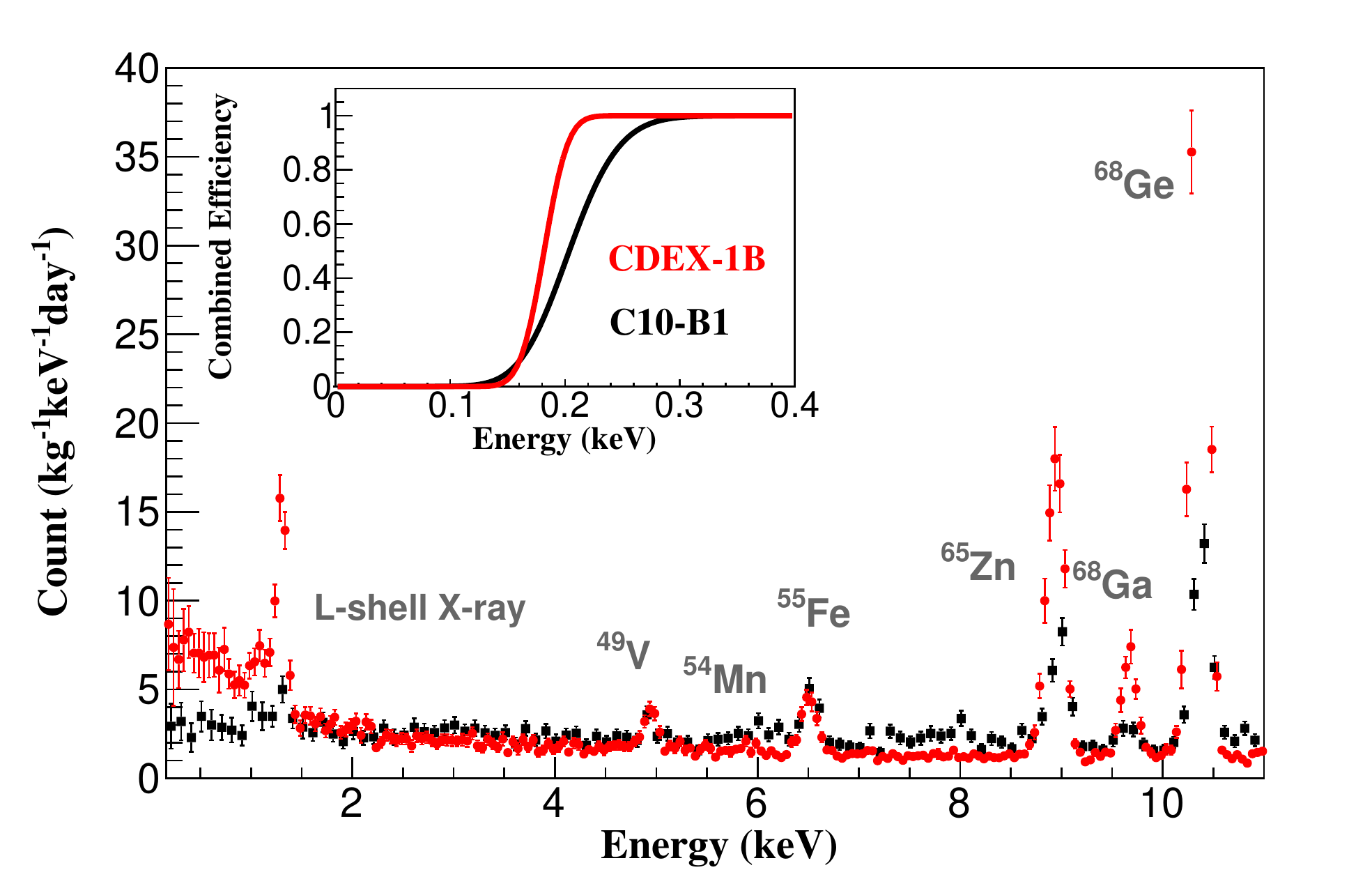}
\caption{Spectra of \textbf{C10-B1 (black points) and CDEX-1B (red points)} after selection. The combined efficiencies are shown in the inset as black line for C10-B1 and as red line for CDEX-1B.}
\label{fig:4}
\end{figure}

As discussed in Ref. \cite{cdex1b2018,cdex10_tech,c10_darkphoton}, the data used in this analysis are obtained through careful selection and the bulk/surface (BS) discriminations \cite{yanglt_bs}. Owing to higher energy threshold and background level, the data from C10-C1 are not taken into consideration in this analysis. The energy spectra of bulk events after all selections, together with the combined selection efficiencies are shown in Fig.~\ref{fig:4} and are based on the exposures of 737.1 kg-days (CDEX-1B) and 205.4 kg-days (C10-B1). The analysis thresholds are both lowered to 160 eV for the CDEX-1B and C10-B1. The uncertainties are mainly from statistics for high energy bins, while the systematic errors originating from the BS discrimination dominate at the low energy range. The details of the BS analysis and uncertainty derivations can be found in Ref. \cite{cdex102018,cdex1b2018}. Different from CDEX-10, the spectrum from CDEX-1B has an anomalous rising profile below 2 keV, which currently cannot be explained well. Several candidate sources (including cosmic rays, neutrons and tritium) were examined but all fail to explain these anomalous events \cite{cdex1b2018}. To further understand this anomaly, more detailed knowledge of material components near the germanium crystal is needed. Research on this issue is ongoing and not within the discussion scope of this article.

\begin{figure}[H]
\centering\includegraphics[width=\linewidth]{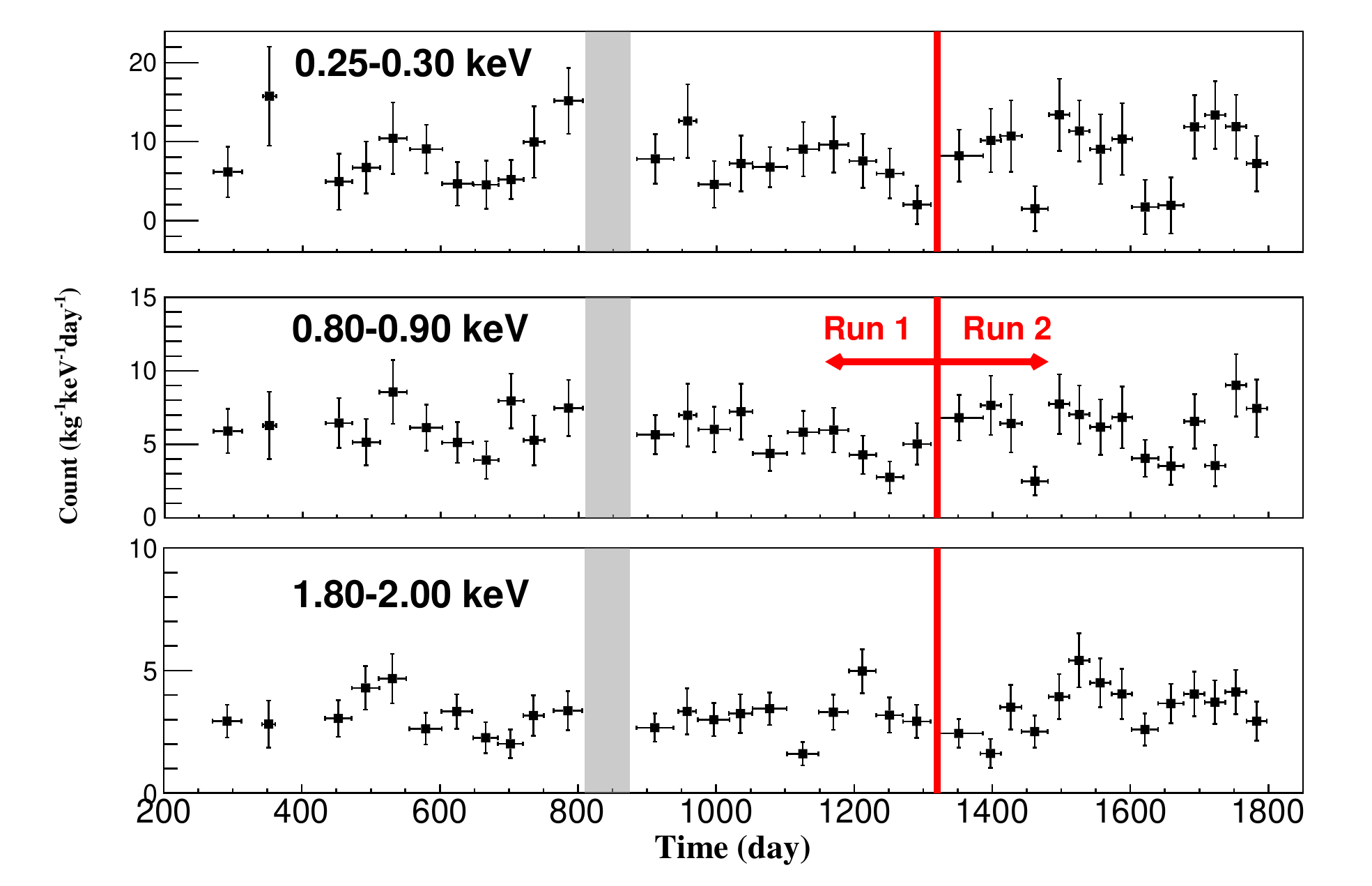}
\caption{Corrected bulk event counts versus time in three different energy bins; bin size selections are consistent with those in data analysis. The horizontal error bars show the time span of each data set. Gray shades represent the time without background data taking owing to neutron and gamma-ray source calibration.}
\label{fig:5}
\end{figure}

The data shown in Fig.~\ref{fig:4} are used in time-integrated (TI) analysis. For AM analysis, the requirement for data is to have stable background with time; CDEX-1B, which has been running stably for approximately 4.2 years, satisfies this requirement. There were two stages during the entire running time of CDEX-1B. Run 1 was from September 27, 2014 until August 2, 2017, when the target was enclosed by an NaI(Tl) anti-Compton detector; Run 2 was from August 4, 2017 until December 2, 2018, when the  NaI(Tl) was replaced by passive copper shielding. Of onte, the total exposure is 1107.5 kg-day within the total time span of 1527 calendar days, and the data used in TI analysis are only derived from Run 1. The data obtained after the selection of three different energy bins are shown in Fig.~\ref{fig:5}.

\section{Result of NREFT}\label{sec:4}
 Both time-integrated (TI) and AM analyses are applied in this analysis using the data from CDEX-1B and CDEX-10. 
 
In TI analysis, the final constraints on the operators $\mathcal{O}_{i}$ are calculated independently and are based on two different datasets from CDEX-1B and CDEX-10 with the exposures of 737.1 kg-day and 205.4 kg-day, respectively. The data used in this analysis were selected based on a series of criteria \cite{cdex1b2018,cdex102018} and the data of CDEX-10 are only from the detector with the best performance, i.e., C10-B1 \cite{c10_darkphoton}.

\begin{figure}[H]
\centering\includegraphics[width=1.0\columnwidth]{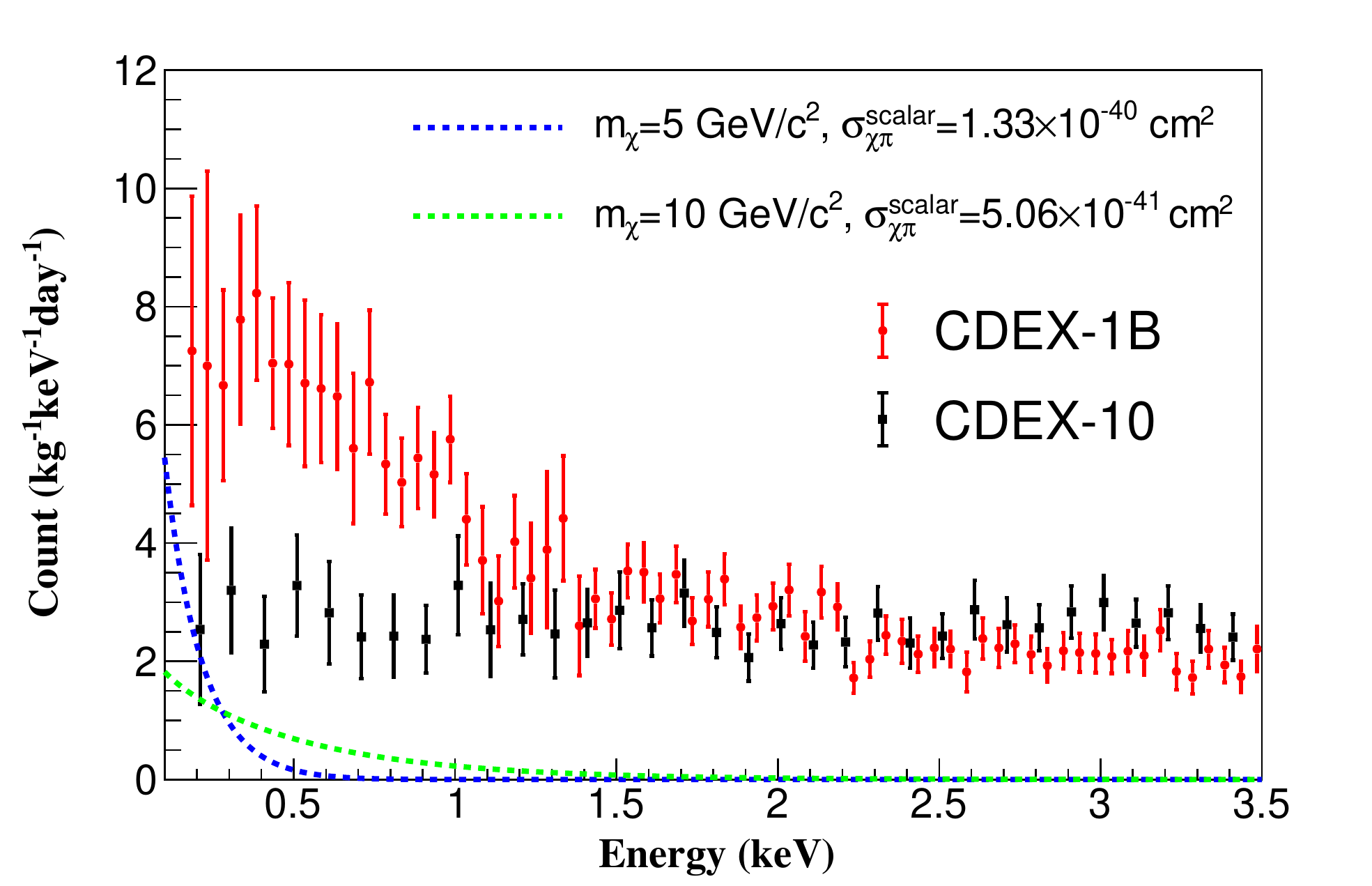}
\caption{Residual spectra of \textbf{C10-B1 (black points) and CDEX-1B (red points)}. Blue dotted and green dotted lines are the expected $\chi$-$\mathcal{N}$ spectra owing to WIMP-pion scattering at $m_{\chi}$ equal to 5 GeV/$c^2$ and 10 GeV/$c^2$ respectively, at cross section corresponding to the upper limits at 90\% C.L., derived from the CDEX-10 dataset \cite{c10_darkphoton}.} 
\label{fig:6}
\end{figure}

\begin{figure}[H]
\centering\includegraphics[width=1.0\columnwidth]{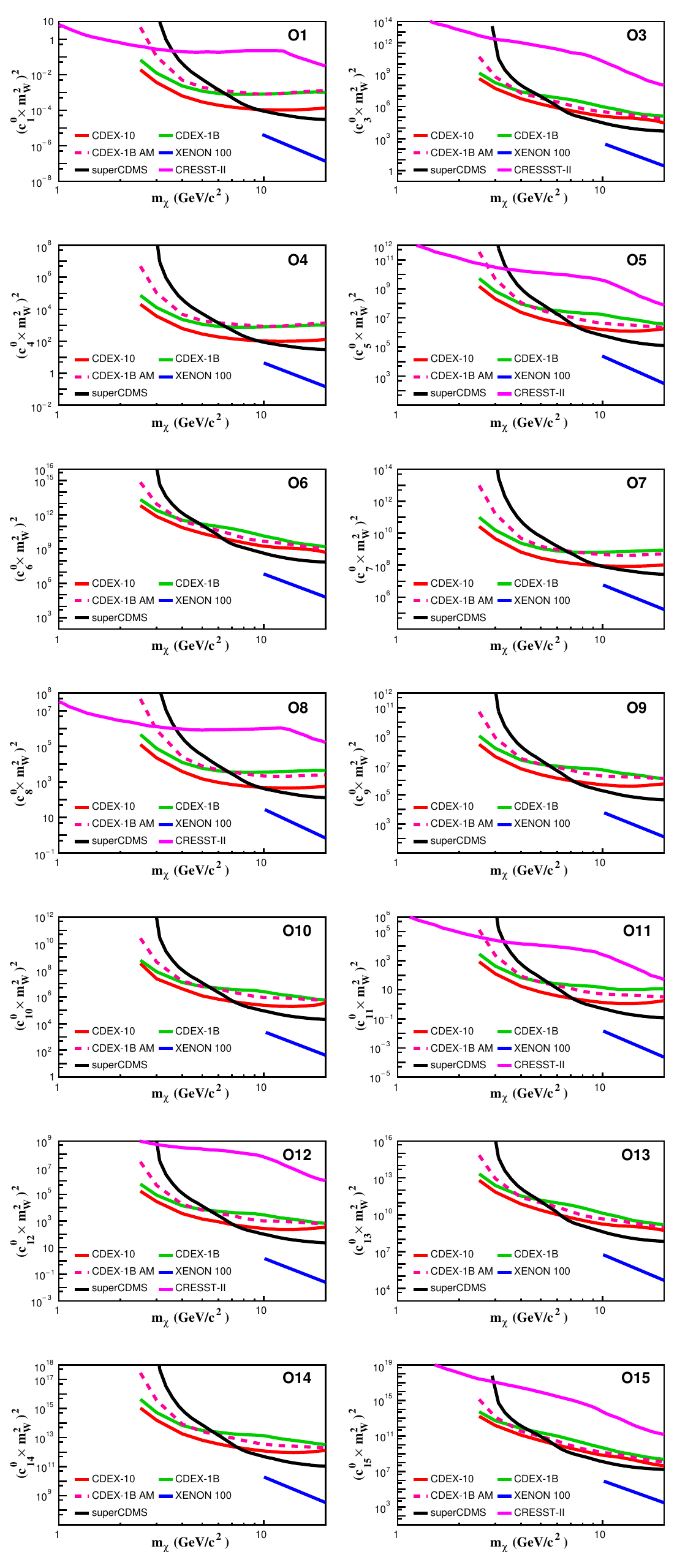}
\caption{The 90\% C.L. upper limits of CDEX-10 and CDEX-1B on isoscalar dimensionless coupling for all NREFT operators. The TI results of CDEX-10 and CDEX-1B are indicated in solid red and green respectively, while the AM results of CDEX-1B are presented as pink dashes. The limits from superCDMS \cite{supercdms_nreft_2015}, XENON100 \cite{xenon_nreft_2017} and CRESST-II \cite{cresst_nreft_2019} are indicated in solid black, blue, and magenta respectively.} 
\label{fig:7}
\end{figure}

For the CDEX-10 data, the minimum-$\chi^2$ analysis \cite{cdex12014} and unified approach \cite{feldman_1998} are applied to the residual spectrum from which the contributions of $L/M$-shell X-ray peaks have been subtracted by fitting corresponding $K$-shell X-ray peaks, as shown in Fig.~\ref{fig:6}. However, for CDEX-1B, the background of unknown origin at low energy region below 2 keV makes the minimum-$\chi^2$ analysis inapplicable and the Binned Poisson method is used as a substitute \cite{binpoisson}. For both CDEX-10 and CDEX-1B, a 10\% systematic error is adopted for the quenching factor calculated by the TRIM package \cite{SOMA_2016,ziegler_2004,linst_2009,scholz_2016}. The upper limits at 90\% confidence level (C.L.) on 14 different operators based on CDEX-1B and CDEX-10 are shown Fig.~\ref{fig:7}, in which the results of SuperCDMS \cite{supercdms_nreft_2015}, XENON100 \cite{xenon_nreft_2017} and CRESST-II \cite{cresst_nreft_2019} are superimposed for comparison. Of note, PandaX-II \cite{pandax_eft_2019} and LUX \cite{lux_nreft} have also released results on EFT studies but with relativistic EFT or pure-proton/neutron framework. The direct comparisons of both results is not feasible. It is observed that CDEX-10 data provide more stringent constraints over the current bounds in the $m_{\chi}$ range from 2.5 up to several GeV$/c^2$ in all operators studied as a consequence of the low physics analysis threshold. 

In the AM analysis, data from CDEX-1B at 0.25$-$5.80 keV are analyzed using the same procedure in Ref \cite{cdex_am}. On request of statistical accuracy in B/S correction, the energy bin sizes are set to be 0.05, 0.1, and 0.2 keV for measured energy at $<$ 0.8, 0.8-1.6, and $>$ 1.6 keV, respectively. The corrected counts of bulk events are denoted by $n_{ijk}$ with $i=$ 1-40 for energy bins, $j=$ 1-35 for time bins, and $k=2$ for Runs 1 and Run 2. Of note, the data in time bins $k=$ 1-21 belong to Run 1, and the rest of the data belong to Run 2. For each energy bin, a minimum-$\chi^{2}$ analysis \cite{cdex_am} is performed simultaneously with 

\begin{equation}
\begin{aligned}
   \chi_{i}^{2}=\sum_{k\in \text{Run}}^{2}\sum_{j\in \text{Time}}^{N}\frac{[n_{ijk}-P_{ijk}-B_{ik}-A_{i}\text{cos}(\frac{2\pi(t_{j}-\Phi)}{T})]^{2}   }{ \Delta_{ijk}^{2}},
\end{aligned}   
\end{equation}
where $\Phi$ and $T$ are the modulation phase and period fixed at 152.5 day and 365.25 day, respectively. $P_{ijk}$ is the time-varying background contributions of the L-shell X-rays from cosmogenic isotopes such as $^{68}$Ge, $^{68}$Ga, and $^{65}$Zn, the intensities of which are fixed by the measured K-shell X-rays at 8.5--10.8 keV. $B_{ik}$ is the time-independent background level and $\Delta_{ijk}^{2}$ are the combined statistical and systematic errors dominated by the B/S correction \cite{yanglt_bs}. $A_{i}$ is the modulation amplitude; it is left unconstrained in the fitting procedure.

Amplitudes $A_i$ are proportional to isoscalar coefficients $(c^{0})^2$ in the form of $A_{i}=(c^{0})^{2}\cdot f(E_{i};\delta E_{i};m_{x})$, where $f$ is a known function of energy $E_{i}$, energy bin size $\delta E_{i}$, and WIMP mass $m_{\chi}$. Thus, the best-fit values of $(c^{0})^{2}$ are then evaluated by minimizing $\sum{\chi_{i}^{2}}$ of Eq. (7), and the upper limits of $(c^{0})^{2}$ at different $m_{\chi}$ are set by unified approach \cite{feldman_1998}. The results of AM analysis are shown in Fig.~\ref{fig:7} together with the results of TI analysis. The operators with a higher order of velocity dependence (all except $\mathcal{O}_{1,4}$) give larger AM amplitudes. Accordingly, as indicated in Fig.~\ref{fig:7}, for a similar CDEX-1B data set, the AM analysis for these operators can give more stringent constraints on coupling coefficients than the TI analysis in the m$_{\chi}$ region above several GeV$/c^2$. This behavior differs from the standard AM/TI analysis in the SI channel \cite{cdex_am}, in which the constraints with TI are more stringent than those of AM over the entire range of $m_{\chi}$.

\section{Results of WIMP-pion scattering in ChEFT }\label{sec:5}

\begin{figure}[H]
\centering\includegraphics[width=1.0\columnwidth]{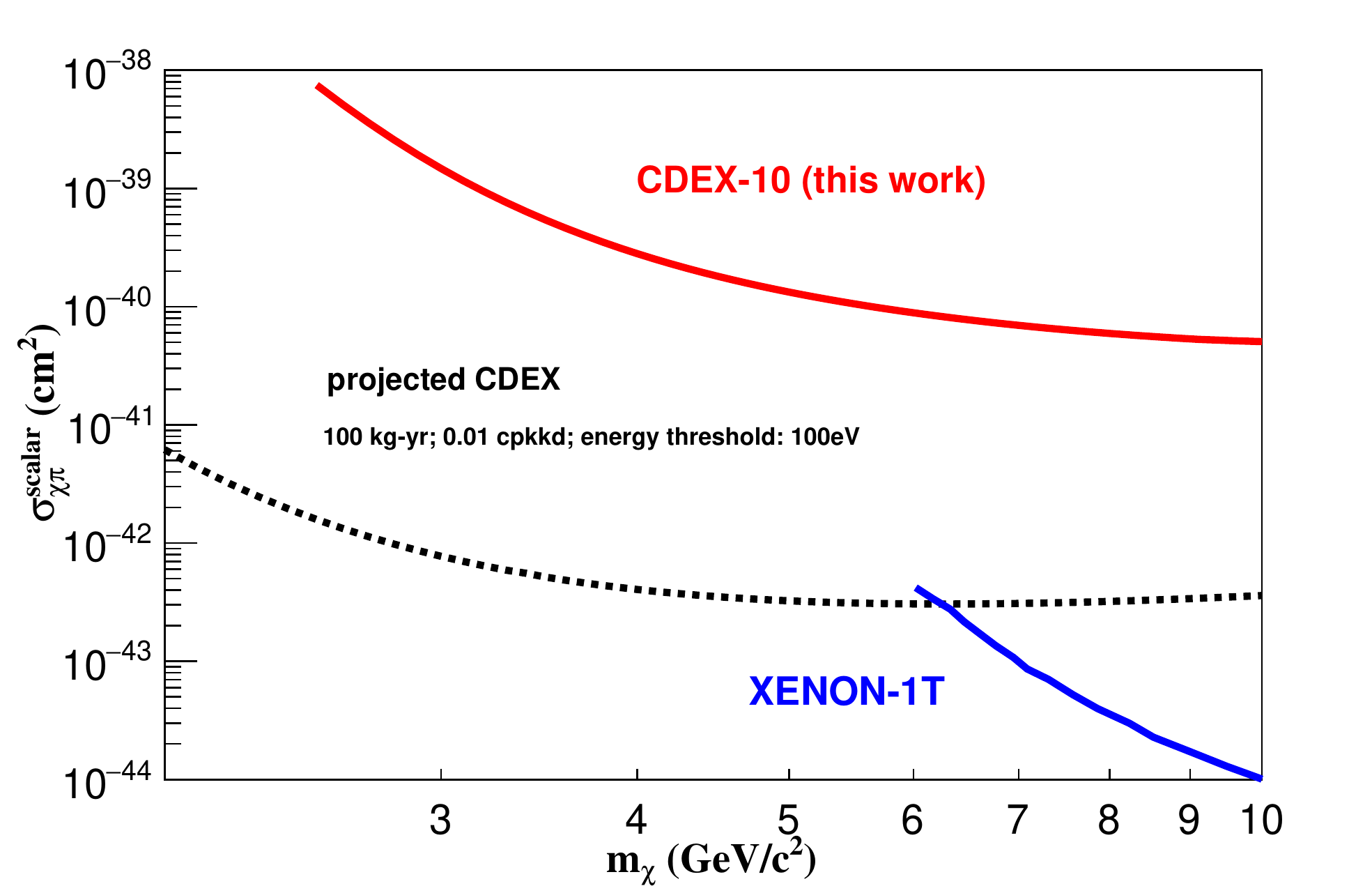}
\caption{Upper limit at 90\% C.L. for the WIMP-pion coupling as a function of WIMP mass for CDEX-10 (red), superimposed with the results (blue) of the XENON1T experiment \cite{xenon_pion}. The potential reach of target sensitivities with a 100-eV threshold at 0.01-counts kg$^{-1}$keV$^{-1}$day$^{-1}$ (cpkkd) background level for 100-kg-year exposure is also shown.} 
\label{fig:8}
\end{figure}

 The WIMP-pion interaction is independent of momentum transfer $\vec{q}$ and velocity $\vec{v}^{\bot}$ such that the AM analysis cannot improve sensitivity. Accordingly, the TI analysis through minimum-$\chi^{2}$ is applied only to the data of CDEX-10, which has lowest background below 2 keV among the CDEX data sets. The fit results at $m_{\chi}=5\:\text{GeV}/c^{2}$ and $m_{\chi}=10\:\text{GeV}/c^{2}$ are shown in Fig.~\ref{fig:6}. The exclusion plot of scalar WIMP-pion coupling at 90\% C.L. is shown in Fig.~\ref{fig:8} and superimposed with the results given by XENON1T \cite{xenon_pion}. The low energy threshold of CDEX-10 produces improved sensitivities for low-mass WIMPs and results in new constraints on the WIMP-pion cross section at $m_{\chi}<$ 6 GeV/$c^2$.

\section{Summary}\label{sec:6}
Because the allowed regions of standard SI and SD are further probed and excluded by various direct detection experiments \cite{cresst_light_2016,supercdms_2018,darkside_2018,xenon_light_2019}, the analyses of new channels, such as NREFT operators and WIMP-pion coupling from ChEFT, are motivating new directions for DM direct detection experiments. By incorporating NREFT into the analysis of the CDEX data, the upper limits on isoscalar coupling are set by both TI and AM analyses at 2.5$-$20 GeV/$c^{2}$ of $m_{\chi}$, and new parameter spaces are excluded at 90\% C.L. For WIMP-pion coupling, CDEX-10 provides new constraints on WIMP-pion cross section at $m_{\chi}<$ 6 GeV/$c^2$. Further improvement in detector performance and background suppression in CDEX-10 and next generation ton-scale experiment \cite{mjl_2019} will provide more stringent on constraints to WIMP coupling in the EFT framework.

This work was supported by the National Key Research and Development Program of China (Grant No. 2017YFA0402200) and the National Natural Science Foundation of China (Grants No. 11725522, No. 11675088, No. 11475099, No. U1865205). We are grateful to Y.F. Zhou for helpful discussion.

\bibliography{cdex_eft}

\end{document}